\documentclass[sigconf,nonacm]{acmart}

\usepackage[nolist]{acronym}
	\begin{acronym}[\qquad\qquad]

\acro{AS}{autonomous system}
	\newacroplural{AS}[ASes]{autonomous systems}
\acro{BCP}{Best Current Practice}
\acro{BGP}{Border Gateway Protocol}
\acro{BYOIP}{bring your own IP addresses}
\acro{CDN}{content distribution network}
\acro{DNS}{Domain Name System}
\acro{DDoS}{distributed denial-of-service}
\acro{DPS}{DDoS protection service}
\acro{DRDoS}{distributed reflection denial-of-service}
\acro{ECMP}{equal-cost multi-path}
\acro{ICMP}{Internet Control Message Protocol}
\acro{IP}{Internet Protocol}
	\acused{IP}
\acro{IPv4}{Internet Protocol version 4}
	\acused{IPv4}
\acro{IPv6}{Internet Protocol version 6}
	\acused{IPv6}
\acro{IRB}{institutional review board}
\acro{ISP}{Internet service provider}
\acro{IX}{Internet exchange}
	\newacroplural{IX}[IXes]{Internet exchanges}
\acro{IXP}{Internet exchange point}
\acro{LDAP}{Lightweight Directory Access Protocol}
\acro{NTP}{Network Time Protocol}
\acro{OS}{operating system}
\acro{pps}{packets per second}
\acro{RSDoS}{ran\-domly-spoofed denial-of-service}
\acro{SAV}{Source Address Validation}
\acro{SOC}{Security Operations Center}
\acro{TTL}{time-to-live}
\acro{UDP}{User Datagram Protocol}

\end{acronym}

\usepackage{algorithm}
\usepackage{algpseudocodex}

\usepackage{booktabs}
\usepackage{graphicx}
\usepackage{multirow}
\usepackage{siunitx}
\usepackage{subcaption}
\usepackage{tikz}
\usepackage{hyperref}

\setcopyright{none}
\DeclareRobustCommand{\circled}[1]{%
	\tikz[baseline=(char.base)]{%
		\node[shape=circle,draw,inner sep=0.5pt] (char) {\small #1};}}

\title{Spoofer or Spoofers? Estimating a Lower Bound on the Number of DRDoS Sources Using Anycast Honeypots}

\author{Bernhard Degen}
\affiliation{%
	\institution{}
	\country{}
	\city{}
}
\author{Bas Palinckx}
\affiliation{%
	\institution{}
	\country{}
	\city{}
}
\author{Mattijs Jonker}
\affiliation{%
	\institution{}
	\country{}
	\city{}
}
\author{Roland van Rijswijk-Deij}
\affiliation{%
	\institution{}
	\country{}
	\city{}
}
\author{Raffaele Sommese}
\affiliation{%
	\institution{}
	\country{}
	\city{}
}

\date{\today}

\begin{document}

\begin{abstract}
	\acs{DDoS} attacks remain a significant threat, with \ac{DRDoS} attacks being particularly difficult to trace back to their sources.
	To better understand attacker behavior and deployment patterns, we present a novel approach for estimating a lower bound on the number of networks involved in generating spoofed traffic.
	Our approach leverages a global deployment of anycast amplification honeypots that attract requests from topologically nearby sources.
	Using this infrastructure, we develop two estimators based on the set of honeypots receiving spoofed traffic and on variations in observed \acs{TTL} values, while accounting for natural path instability.
	Analyzing 287~days of amplification attacks, we find that at least 21.0\% originate from multiple network locations, indicating that attackers frequently distribute spoofing activity across networks.
	Our findings suggest that combating spoofing requires coordinated and distributed defenses, and inform the design of future attribution techniques.
\end{abstract}

\maketitle

\section{Introduction}%
\label{sec:introduction}

\Ac{DDoS} attacks are a persistent threat, with 14.6\,M~\ac{IP} addresses targeted between 2019 and mid-2023~\cite{HiesgenEtAl_AgeDDoScoveryEmpirical_2024}.
Approximately half of these were \acf{DRDoS} attacks.
In such attacks, an attacker spoofs the source \ac{IP} address of a packet to amplify their attack capacity while obfuscating the true network location.
A result of the bandwidth amplification in these types of attacks~\cite{Rossow_AmplificationHellRevisiting_2014} is that an attacker needs only limited resources to generate a substantial volume of malicious traffic.
This asymmetry between the input (small request) and output (large response) enables even single hosts to mount a disruptive attack.

These attacks are possible because approximately one quarter of \acp{AS} do not fully implement outbound \ac{SAV}~\cite{CenterforAppliedInternetDataAnalysisCAIDA_Spoofer_2026}, making their infrastructure attractive to attackers for launching \ac{DRDoS} attacks.
\Ac{DRDoS} attacks have been extensively studied, but there has been limited focus on finding spoofing sources and the infrastructure they rely on.
Identifying the spoofing sources of a \ac{DDoS} attack is challenging for several reasons.
First, the source addresses of \ac{DRDoS} traffic are spoofed, which obscures the attacker's identity.
Second, neither \ac{IPv4} nor \ac{IPv6} exposes information about the path a packet has taken through the network.
By design, routers forward packets while modifying only a limited set of header fields.
As a result, reconstructing a packet's path requires extensive cooperation among network operators across multiple administrative and jurisdictional boundaries, which is rarely feasible in practice.

In this study, we develop a methodology to estimate a lower bound on the number of spoofing sources involved in an attack.
Our approach leverages a network of anycast amplification honeypots, deployed across six continents, that mimic reflectors (e.g.,~open resolvers) to capture \ac{DRDoS} activity.
Using \ac{TTL} values of incoming attack traffic, we derive a lower bound on the number of spoofing sources.
We deliberately choose to estimate a lower bound as our approach cannot distinguish between spoofing sources with identical path lengths toward our honeypots.
We provide a detailed explanation of the intuition behind our approach and how we model path variability using RIPE Atlas in~\S\ref{sec:measuring_ttl_stability} and~\S\ref{sec:computing_an_interval_cover}.
Building on this approach, our contributions are as follows:

\begin{enumerate}
	\item We leverage a 32-node, globally distributed anycast testbed to deploy amplification honeypots that provide a location-aware view of attack activity at Internet scale.
		With 287~days of honeypot data across two distinct periods, we assess the repeatability of our approach.
	\item We establish a baseline of \ac{TTL} variability using widespread probing from \num{12502} Atlas probes.
	\item We propose and evaluate two estimators that establish lower bounds on the number of spoofing sources, calibrated using empirically derived thresholds, and show that \ac{TTL} frequencies can further tighten these bounds.
	\item We show that at least 21.0\% of attacks involve multiple spoofing sources, highlighting the need for coordinated and distributed defenses.
\end{enumerate}

We make the \ac{TTL} measurements performed in this study available for reproducibility and further analysis.

\section{Background}%
\label{sec:background}

In this section, we provide background on amplification attacks (\S\ref{sec:amplification_attacks}) and explain how the \ac{IP} \ac{TTL} field can be used to infer the number of hops a packet has traversed~(\S\ref{sec:inferring_hop_counts}).

\subsection{Amplification attacks}%
\label{sec:amplification_attacks}

In a \ac{DDoS} attack, a system is targeted by a high volume of requests with the aim of rendering it unavailable.
In this work, we focus on \ac{DRDoS} attacks, in which, unlike direct-path attacks, the traffic is reflected via an intermediate host (the reflector).
With this type of attack, the attacker forges the source address of requests by setting it to that of the target, causing the intermediate host to send replies to the target.
In a volumetric attack, the attacker aims to solicit responses that are larger than the request, thereby amplifying the attack power.
This technique is typically performed against protocols amenable to amplification (i.e.,~sending a response exceeding the request in size), such as the \acf{DNS} and \acf{NTP}.

Before an attacker can launch an attack, it must first identify servers that can be exploited as amplifiers.
This is commonly done by scanning the Internet for services vulnerable to amplification.
The resulting list can be reused to launch multiple attacks~\cite{KruppEtAl_IdentifyingScanAttack_2016}.
Amplification honeypots can observe both the scanning and the subsequent spoofed requests used in attacks.
In~\S\ref{sec:monitoring_attacks}, we describe how we use honeypots to collect attacks.

\subsection{Inferring hop counts}%
\label{sec:inferring_hop_counts}

Attackers may send spoofed traffic from multiple network locations to increase attack power or evade filtering mechanisms.
Because the source addresses in amplification requests are spoofed, we must rely on other signals to distinguish between these spoofing sources.
Therefore, we estimate the number of hops that packets traversed from the \ac{TTL} field of the \ac{IPv4} header.
The \ac{TTL} field prevents routing loops: each router along the path decrements the value by one, and once it reaches zero the packet is discarded.
\Acp{OS} typically initialize the \ac{TTL} with a default of 64, 128, or 255~\cite{Zalewski_P0fREADME_2012}.
Because these initial values are spaced far apart relative to typical Internet path lengths, the hop count can be estimated by subtracting the observed \ac{TTL} from the closest higher initial value.
Consequently, the \ac{TTL} field provides an indication of the path length from the sender to the receiver.
While an attacker can choose the initial \ac{TTL} value, it cannot prevent its subsequent hop-wise decrements.
Attacks whose initial \ac{TTL} value markedly deviates from the default \ac{OS} values are straightforward to detect, whereas per-request \ac{TTL} manipulation introduces noise in our data~(\S\ref{sec:sensitivity_to_ttl_manipulation}).
Furthermore, Internet paths are inherently dynamic, and routing changes, load balancing, and other network device behavior can cause natural path instability~(\S\ref{sec:measuring_ttl_stability}).

\section{Related work}%
\label{sec:related_work}

We begin by describing existing work on observing and characterizing \ac{DDoS} attacks~(\S\ref{sec:characterizing_ddos_attacks}).
We then survey research that investigates how spoofed traffic can be identified~(\S\ref{sec:identifying_spoofed_traffic}) and traced to its sources~(\S\ref{sec:attributing_spoofed_traffic}).

\subsection{Characterizing DDoS attacks}%
\label{sec:characterizing_ddos_attacks}

\Ac{DDoS} activity can be observed using different vantage points.
Back\-scatter traffic observed at network telescopes allows inferring \ac{RSDoS} attacks~\cite{MooreEtAl_InferringInternetDenialofService_2006}.
If the attacker is generating source addresses uniformly at random, the target's replies can be observed at any telescope of sufficient size.

However, in \ac{DRDoS} attacks, the attacker uses the address of their target~\cite{Paxson_AnalysisUsingReflectors_2001}.
Therefore, a network telescope does not receive replies, necessitating a different approach.
Instead, researchers often rely on amplification honeypots to observe this type of attack~\cite{%
	KramerEtAl_AmpPotMonitoringDefending_2015,
	KruppEtAl_IdentifyingScanAttack_2016,
	NoroozianEtAl_WhoGetsBoot_2016,
	ThomasEtAl_1000DaysUDP_2017,
	GriffioenEtAl_ScanTestExecute_2021,
	HeinrichEtAl_NewKidsDRDoS_2021,
	NawrockiEtAl_FarSideDNS_2021}.

Alternatively, \ac{DRDoS} attacks can be observed from traffic traces collected at \acp{IXP}~\cite{KoppEtAl_DDoSNeverDies_2021, NawrockiEtAl_FarSideDNS_2021}, and from \ac{BGP} blackholing activity~\cite{GiotsasEtAl_InferringBGPBlackholing_2017}.
However, Hiesgen et al.~\cite{HiesgenEtAl_AgeDDoScoveryEmpirical_2024} noted considerable differences in long-term attack trends and targets reported by academic and industrial data sources.
Leveraging a novel anycast honeypot deployment, we provide new insights into the infrastructure potentially involved in generating spoofed traffic for \ac{DRDoS} attacks.

\subsection{Identifying spoofed traffic}%
\label{sec:identifying_spoofed_traffic}

Spoofed traffic can be filtered at end-hosts by learning a mapping of valid \ac{TTL} values for each source address and filtering traffic that deviates too much from the expected value~\cite{JinEtAl_HopCountFilteringEffective_2003, MukaddamEtAl_IPSpoofingDetection_2014}.
However, false positives risk filtering legitimate traffic, and an attacker might infer the expected \ac{TTL} value to circumvent this protection~\cite{BackesEtAl_FeasibilityTTLBasedFiltering_2016}.
Other approaches detect spoofing at \acp{IXP}~\cite{LichtblauEtAl_DetectionClassificationAnalysis_2017} and \acp{ISP}~\cite{GigisEtAl_BadPacketsCome_2024}.
More closely related, active anycast catchment mappings can be used to derive for each catchment a list of prefixes from which traffic may originate~\cite{deVriesEtAl_GlobalScaleAnycastNetwork_2020}.
Any traffic observed outside these catchments would be flagged as likely spoofed.
Rather than using \ac{TTL} values for spoofing detection alone, our approach leverages them to distinguish multiple spoofing sources within a catchment using anycast-hosted honeypots, enabling finer-grained visibility into spoofing behavior beyond traditional unicast honeypot deployments.

\subsection{Attributing spoofed traffic}%
\label{sec:attributing_spoofed_traffic}

Several techniques have been proposed in which on-path routers
	mark packets~\cite{%
		SavageEtAl_PracticalNetworkSupport_2000,
		SongPerrig_AdvancedAuthenticatedMarking_2001,
		BelenkyAnsari_IPTracebackDeterministic_2003,
		YaarEtAl_PiPathIdentification_2003%
	},
	report them to the destination via diagnostic messages~\cite{%
		WuMassey_IntentionDrivenICMPTraceBack_2001,
		BellovinEtAl_ICMPTracebackMessages_2003,
		LeeEtAl_ICMPTracebackCumulative_2003},
	log them~\cite{%
		DuffieldGrossglauser_TrajectorySamplingDirect_2000,
		SnoerenEtAl_HashBasedIPTraceback_2001,
		SungEtAl_LargeScaleIPTraceback_2008},
	construct an overlay network~\cite{%
		Stone_CenterTrackIPOverlay_2000,
		CastelucioEtAl_ASlevelOverlayNetwork_2009,
		CastelucioEtAl_IntradomainIPTraceback_2012},
	or use a combination thereof to reconstruct their path.
However, all of these approaches require wide-scale router support and have not seen any notable adoption.
Other ideas include link flooding~\cite{BurchCheswick_TracingAnonymousPackets_2000}, collecting \ac{ICMP} packets generated by on-path devices~\cite{YaoEtAl_PassiveIPTraceback_2015} to reconstruct the path to the spoofing source, selectively responding to addresses to link scanners to spoofers~\cite{KruppEtAl_IdentifyingScanAttack_2016}, and training machine learning models on self-attacks~\cite{KruppEtAl_LinkingAmplificationDDoS_2017}.

Furthermore, it is possible to reduce the set of possible spoofing sources using \ac{BGP} traffic engineering techniques.
Fonseca et al.~\cite{FonsecaEtAl_TrackingSourcesSpoofed_2020} proposed using anycast, \ac{BGP} path prepending and poisoning to reduce the number of \acp{AS} that can originate spoofed traffic.
Using 705~announcement configurations (taking an estimated 34.3~days), their method partitioned \acp{AS} into increasingly smaller sets, 92\% of which contain only one \ac{AS}.
Krupp and Rossow~\cite{KruppRossow_BGPeekaBooActiveBGPbased_2021} introduced an approach in which \acp{AS} are systematically poisoned.
Using \ac{AS} relationships, their approach requires a median time of 16.4~hours, rendering it impractical for tracing real-world attacks, which have a median duration of 15~minutes or less~\cite{KramerEtAl_AmpPotMonitoringDefending_2015, JonkerEtAl_MillionsTargetsAttack_2017, ThomasEtAl_1000DaysUDP_2017, GriffioenEtAl_ScanTestExecute_2021}.
Our approach enables real-time operation: it can be initiated as soon as an attack is observed and provides an immediate lower-bound estimate of the number of spoofing sources.
Additionally, it does not require changes to protocols or on-path devices.

\section{Methodology}%
\label{sec:methodology}

In this section, we describe the steps to arrive at a lower bound estimate of the number of spoofing sources.
We first describe how the \ac{IP} \ac{TTL} field can be used to infer the number of hops a packet has traversed~(\S\ref{sec:inferring_hop_counts}).
Next, we explain how we use anycast honeypots to measure \ac{DRDoS} attack activity~(\S\ref{sec:monitoring_attacks}).
We then describe how we measure the expected variability in \ac{TTL} values observed at each honeypot~(\S\ref{sec:measuring_ttl_stability}).
This information allows us to establish thresholds that distinguish multiple spoofing sources from natural path instability.
Finally, using these thresholds, we introduce two estimators that provide a lower bound on the number of sources per honeypot~(\S\ref{sec:computing_an_interval_cover}).

\subsection{Monitoring attacks}%
\label{sec:monitoring_attacks}

Our measurement setup consists of an anycast testbed comprising 32 geographically distributed nodes across six continents.
The testbed is hosted on Vultr, a cloud provider supporting \ac{BYOIP} and \ac{BGP} sessions~\cite{Vultr_32CloudData_2026}.
Through Vultr's \ac{AS}, 30 of the 32 nodes have connectivity to at least one \ac{IX} in addition to upstream transit providers.
Each node is assigned both a unicast address and a shared anycast address that is announced by all nodes.
The global presence of the testbed nodes helps attract localized attack activity.
As discussed in~\S\ref{sec:amplification_attacks}, in a \ac{DRDoS} attack the attacker spoofs requests to a service amenable to amplification in an attempt to direct a larger response to their target.
To observe such attacks, we deployed AmpPot~\cite{KramerEtAl_AmpPotMonitoringDefending_2015}, a honeypot that emulates an amplifier.
AmpPot exposes 16~amplifiable \ac{UDP} protocols%
	\footnote{CHARGEN, DNS, Jenkins, LDAP, MSSQL, Memcached, NetBIOS, NTP, QOTD, RIPv1, RPC, SIP, SIPS, SNMP, SSDP, and TFTP.}
	and captures requests to their respective ports.
Since our focus is on estimating the number of spoofing sources regardless of protocol, we include all protocols in our analysis.
We configured AmpPot to listen for requests on both the unicast and anycast addresses to maximize attack coverage.

\subsubsection{Rate limiting and sampling}%
\label{sec:rate_limiting_and_sampling}

As our honeypots pose as amplifiers, it is imperative to prevent any harm resulting from our responses.
We follow best practices and apply the default rate limit for AmpPot.
Appendix~\ref{sec:amplification_honeypots} details the precautions implemented to minimize any potential harm from our honeypot deployment.

Besides limiting requests, AmpPot applies sampling to the responses it records to avoid saturating storage.
It records the first 100~requests per honeypot and per /24~network.
After that, requests are sampled at a rate of~$\frac{1}{100}$ until the arrival rate between two subsequent packets from the /24~network exceeds 60~seconds.
This sampling may affect the precision of estimating the scale of attacks~(\S\ref{sec:attack_scale_and_spoofing_sources}).

\subsubsection{Attack grouping}%
\label{sec:attack_grouping}

AmpPot logs individual requests, which result from scans, attacks, or other activity.
To infer attacks from these requests, we followed the AmpPot attack definition~\cite{KramerEtAl_AmpPotMonitoringDefending_2015, KruppEtAl_IdentifyingScanAttack_2016}:
	a sequence of at least 100~packets, each received no more than one hour apart, with a single source \ac{IP}~address using a single application-layer protocol to any honeypot \ac{IP}~addresses.
This definition has two implications.

First, the number of inferred attacks increases as more honeypots are monitored, since a single attack may reach multiple honeypots, making it more likely to meet the threshold.
We therefore use both anycast and unicast requests to increase coverage.
However, we only perform spoofed source count estimation on requests reaching the anycast address, since anycast distributes traffic across multiple sites, which provides additional information that improves estimation accuracy~(\S\ref{sec:computing_an_interval_cover}).

Second, the definition treats multiple bursts exceeding the 100-packet threshold that occur less than one hour apart as a single sustained attack.
Conversely, amplification requests targeting the same host but using different protocols are considered separate attacks.

\subsection{Measuring TTL stability}%
\label{sec:measuring_ttl_stability}

Our goal is to infer a lower bound on the number of spoofing sources from amplification requests observed at honeypots.
We define a subnet sending spoofed traffic as a \emph{spoofing source}, and in our analysis, multiple hosts sending spoofed packets from the same subnet are considered the same spoofing source, as their path lengths are likely identical.
As explained in~\S\ref{sec:inferring_hop_counts}, we infer path length from the \ac{TTL} value included in each packet.

In practice, Internet paths can fluctuate in length, causing variations in the observed \ac{TTL} values even from a single spoofing source.
Distinct \ac{TTL} values can serve as a proxy for spoofing sources, but naively counting them at each honeypot would likely overestimate their number because of these fluctuations.
Therefore, we allow some variation in \ac{TTL} values based on the expected variability in \ac{TTL} values before considering a new spoofing source.
\Ac{TTL} variability differs per honeypot because the hosting anycast site is connected to different upstream networks and has a unique catchment.
To obtain a lower bound on the number of spoofing sources in an attack, we performed a calibration step to derive a per-honeypot \ac{TTL} stability threshold~$j_h$ based on the measured \ac{TTL} jitter at honeypot~$h$.

As mentioned earlier, we aim to estimate a lower bound on the number of spoofing sources.
The limited range of Internet path hop counts makes it impossible to determine whether observing only a single distinct \ac{TTL} value across multiple packets indicates a single spoofing source or multiple spoofing sources.
Hence, we cannot control for this confounding factor.
However, overestimation can be constrained by learning the expected \ac{TTL} variability per honeypot.
By favoring underestimation, our estimator remains conservative: when it indicates multiple spoofing sources, this conclusion is well supported.

To establish a TTL stability threshold $j_h$, we perform a measurement to empirically estimate $j_h$ for all of our honeypots.
To ensure a representative estimation, we need to probe our sites from a topologically diverse set of locations.
We make use of the RIPE Atlas~\cite{RIPENetworkCoordinationCenter_RIPEAtlas_2025} platform to achieve this.
Atlas consists of geographically distributed volunteer-operated probes to perform measurements.
To maximize coverage, we requested the maximum number of Atlas probes available to us~(\num{15000}), randomly distributed worldwide.
From each, we sent six \ac{DNS} queries to an address within the honeypots' /24~anycast prefix every three hours for 24~hours.
The six queries are spaced five seconds apart, and the start times are uniformly distributed within the three-hour interval.
Even though 98.9\% of the attacks we observed lasted less than 24~hours, we ran the experiment for a full day to capture diurnal patterns.

A potential source of instability in observed path lengths is variation in packet header fields used for load balancing.
We embed the probe identifier and timestamp in the query name and rely on RIPE Atlas's default behavior of randomizing the source port for each query.
Since \ac{ECMP} implementations commonly perform load balancing based on a hash of the flow 5-tuple~\cite{HendriksEtAl_LoadBalancingAnycastFirst_2026}, which includes the source port, queries are more likely to be distributed across multiple paths in this configuration where random source ports are used.
We acknowledge that explicitly triggering load balancing in our calibration experiment imposes a limitation, as it carries the implicit assumption that attackers also randomize source ports and thus trigger load balancing.
However, in 23.7\% of the attacks we observed, a static source port is used, which may make our lower bound more conservative than necessary in these cases.

Let $A$ be the set of Atlas probes.
For each Atlas probe $a \in A$ and honeypot $h$, we calculate the range of \ac{TTL} values
	$$ \delta_{h,a} = \max(T_{h,a}) - \min(T_{h,a}), $$
	where $T_{h,a}$ denotes the observed \ac{TTL} values from probe $a$ to honeypot $h$.
Next, we compute the per-honeypot stability threshold $j_h$ as half of the 95\textsuperscript{th} percentile of the \ac{TTL} ranges observed from probes reaching honeypot~$h$, rounded up:
	$$ j_h = \left\lceil P_{95} \big( \{ \delta_{h,a} \mid a \in A \} \big) \, / \, 2 \right\rceil. $$

The division by two accounts for the fact that \ac{TTL} deviations may occur in both directions relative to a typical observed value.
We use the 95\textsuperscript{th} percentile to capture the expected \ac{TTL} variability across all probes, while reducing the influence of outliers.
This presents a trade-off: a higher percentile makes the method more robust to outliers as these larger deviations are more likely to be included in the threshold; a lower percentile makes it more sensitive to natural path instability.

We do not model variability in hop counts as a stochastic process because hop counts reflect deterministic routing decisions (such as \ac{ECMP} routing) and occasional path changes or route flapping, rather than continuous random variability.
Although variability resulting from \ac{ECMP} could, in principle, be modeled based on the packet header fields observed by the honeypots, doing so is beyond the scope of this work.

After performing the measurements, we observed 114~probes with \ac{TTL} values higher than the expected initial value (64), likely due to middleboxes rewriting \ac{TTL} values.
Of these, 11 had inconsistent inferred initial \ac{TTL} values, which we discarded to avoid overestimating variability, leaving \num{12502}~probes for analysis.
Applying our per-honeypot threshold method, the distribution of thresholds~$j_h$ across the 32~honeypots was as follows:
	most honeypots (26 out of 32) had $j_h = 1$, indicating that the expected \ac{TTL} variability is generally very small.
Three honeypots had $j_h = 0$, indicating nearly no variability, while the remaining three had $j_h = 2$, reflecting higher variability.

Figure~\ref{fig:ttl_range_distr} (Appendix~\ref{sec:detailed_calibration_results}) provides detailed per-honeypot \ac{TTL} range distributions.
It shows substantial differences across honeypots, supporting the use of per-honeypot thresholds.
It also shows a small number of outliers exceeding the 95\textsuperscript{th} percentile, which do not influence the resulting thresholds.

One limitation is that anycast catchments can differ substantially in size~\cite{DegenEtAl_EmpiricalCharacterizationAnycast_2024,HendriksEtAl_EmpiricalEvaluationLongitudinal_2025}.
While we selected the maximum number of Atlas probes available to us, four honeypots had fewer than 50~Atlas nodes routing to them.
Due to the limited diversity, the derived \ac{TTL} stability thresholds for these honeypots may be less robust.

As a validation step, we repeated the calibration experiment 99~days after the original experiment, and observed similar results: two honeypots changed from $j_h = 1$ to $j_h = 2$, while one honeypot changed from $j_h = 2$ to $j_h = 1$.
Overall, the thresholds remained stable despite potential routing and topological changes, indicating that the measured \ac{TTL} variability is relatively consistent over time.

\subsection{Computing an interval cover}%
\label{sec:computing_an_interval_cover}

Since an attacker cannot control which anycast site their traffic is routed to, traffic from multiple spoofing sources may reach different sites.
However, when multiple spoofing sources are topologically close, they may still be routed to the same site.
Consequently, the number of sites receiving spoofed attack traffic provides a loose lower bound on the number of spoofing sources active.
We call this baseline heuristic \emph{honeypots hit} hereafter.

Although sources within a single network generally have affinity to a single anycast site, a small fraction of source /24~prefixes~(4.4\%) are observed to be routed to multiple sites~\cite{HendriksEtAl_LoadBalancingAnycastFirst_2026}.
Because we treat traffic arriving at different sites as coming from separate spoofing sources, this routing instability can lead us to slightly overestimate the number of spoofing sources.
However, previous work has shown that short-term catchment stability is high; 99\% of /24~prefixes route to the same site from one day to the next~\cite{HendriksEtAl_EmpiricalEvaluationLongitudinal_2025}.

The distribution of \ac{TTL} values at each honeypot provides an additional indicator of the number of spoofing sources \emph{per honeypot}.
Intuitively, when the observed \ac{TTL} values at a honeypot are far apart, it suggests that traffic is arriving from multiple sources located in networks with different path lengths to the honeypot.
Based on this intuition, we estimate a lower bound on the number of spoofing sources per honeypot using the observed \ac{TTL} values, and sum these estimates across honeypots to obtain the final lower bound.
We propose two estimators: one that gives an absolute lower bound on the number of spoofing sources, the minimum interval cover, and one that may more realistically reflect traffic patterns, the frequency-weighted~interval cover.
The remainder of this section details these estimators.

\subsubsection{Minimum interval cover}%
\label{sec:minimum_interval_cover}

\begin{algorithm}[tpb]
	\caption{Minimum interval cover}%
	\label{alg:ttl_min_cover}
	\begin{algorithmic}[1]
		\Require TTL values $T = \{t_1, t_2, \dots, t_n\}$, threshold $j$
		\Ensure interval centers $C$
		\State sort $T$ in ascending order
		\State $C \gets \emptyset$
		\State $i \gets 1$
		\While{$i \le n$}
			\State $c \gets T[i] + j$
			\State $C \gets C \cup \{c\}$
			\While{$i \le n \land T[i] \le c + j$}
				\State $i \gets i + 1$
			\EndWhile
		\EndWhile
		\State \Return $C$
	\end{algorithmic}
\end{algorithm}

We use a greedy algorithm~\cite[Chapter 16]{CormenEtAl_IntroductionAlgorithms_2009} to compute an unweighted interval cover.
For each honeypot, given a set of observed \ac{TTL} values~$T$ and a threshold~$j$, we compute the smallest set of interval centers~$C$ such that for each $c \in C$, $c \pm j$ covers all $t \in T$.
As shown in Algorithm~\ref{alg:ttl_min_cover}, we compute $C$ by first sorting the \ac{TTL} values~$T$ in ascending order and iteratively placing a new interval with its lower bound equal to the smallest uncovered value.
This greedy strategy is optimal for the minimum interval cover problem because each interval covers the lowest available point while extending its reach as far as possible.
Therefore, it provides a lower bound under the assumption all \ac{TTL} values from each spoofing source vary by at most $j$.

\subsubsection{Frequency-weighted interval cover}%
\label{sec:frequency_weighted_interval_cover}

\begin{algorithm}[tpb]
	\caption{Frequency-weighted interval cover}%
	\label{alg:ttl_freq_cover}
	\begin{algorithmic}[1]
		\Require TTL values $T = \{t_1, t_2, \dots, t_n\}$, frequency map $f \colon T \to \mathbb{N}$, threshold $j$
		\Ensure interval centers $C$
		\State $\text{covered}[t] \gets \textbf{false}$ for all $t \in T$
		\State $C \gets \emptyset$
		\While{$\exists\, t \in T \text{ such that covered}[t] = \textbf{false}$}
			\State $c \gets \arg\max_{t \colon \text{covered}[t] = \textbf{false}} f(t)$
			\State $C \gets C \cup \{c\}$
			\ForAll{$t \in T$}
				\If{$\text{covered}[t] = \textbf{false} \land \left|\, t - c \,\right| \le j$}
					\State $\text{covered}[t] \gets \textbf{true}$
				\EndIf
			\EndFor
		\EndWhile
		\State \Return $C$
	\end{algorithmic}
\end{algorithm}

While the minimum interval cover provides a minimal solution, it may underestimate the number of spoofing sources in practice because the interval centers are not necessarily aligned with the most frequent values.
To address this, we introduce a frequency-weighted interval cover that accounts for the observed frequencies of \ac{TTL} values.
Algorithm~\ref{alg:ttl_freq_cover} formalizes this approach: at each step, it selects the uncovered \ac{TTL} with the highest frequency as an interval center and marks all \ac{TTL} values within the interval as covered.
This process continues until all \ac{TTL} values are covered.
In contrast to the minimum interval cover, the result of the frequency-based interval cover may yield overlapping intervals, which signify \ac{TTL} observations that could have been sent by either inferred source identified by the surrounding interval centers.

\begin{figure}[tpb]
	\centering
	\begin{subfigure}[t]{.5\linewidth}
		\centering
		\captionsetup{width=.9\linewidth}
		\includegraphics[width=\linewidth]{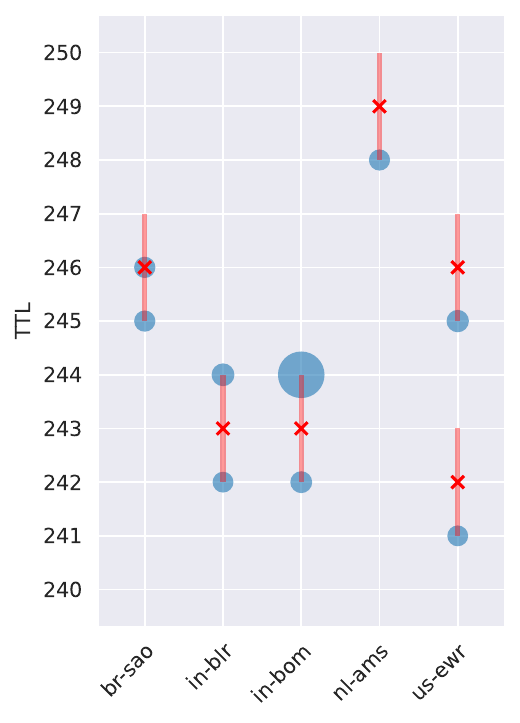}
		\Description{}
		\caption{Minimum interval cover (inferring 6~sources)}%
		\label{fig:ttl_cover_min}
	\end{subfigure}%
	\hfil
	\begin{subfigure}[t]{.5\linewidth}
		\centering
		\captionsetup{width=.9\linewidth}
		\includegraphics[width=\linewidth]{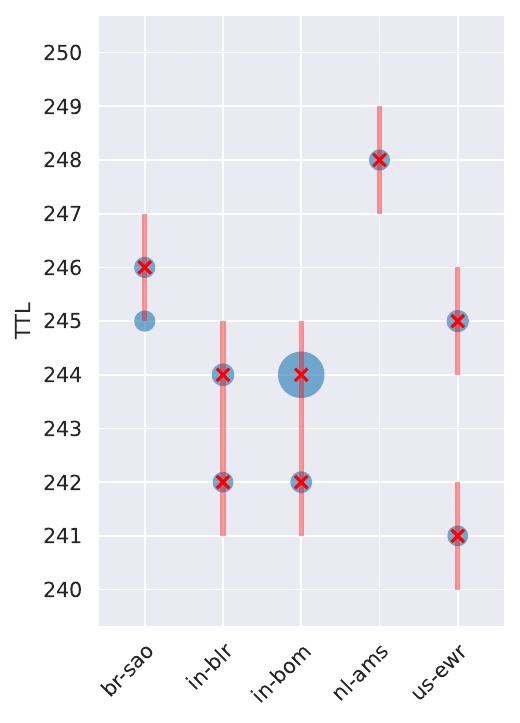}
		\Description{}
		\caption{Frequency-weighted interval cover (inferring 8~sources)}%
		\label{fig:ttl_cover_freq}
	\end{subfigure}%
	\caption{%
		\acs{TTL} covering algorithms applied to \acs{TTL} values observed per honeypot $h$ ($x$-axis).
		The size of a circle indicates the frequency of the observation.
		The bars indicate the allowed variation $j_h$ (here 1 for all $h$) relative to the inferred cover center $c$~(cross).
	}
	\label{fig:ttl_cover}
\end{figure}

Figure~\ref{fig:ttl_cover} illustrates both algorithms using an \acs{NTP} attack that we observed targeting a host in France.
The minimum interval cover starts intervals at the lowest uncovered \ac{TTL} values.
In contrast, the frequency-weighted interval cover aligns the interval centers with the uncovered \ac{TTL} values ordered by frequency per honeypot, inferring one additional spoofing source at both \texttt{in-blr} and \texttt{in-bom}.
This occurs because the minimum interval cover selects the lowest observed \ac{TTL} value~(242) as the start of the first interval for both honeypots, thereby covering all observations, whereas the frequency-weighted interval cover selects the most frequently observed \ac{TTL} value~(244 in both cases), requiring an additional interval to cover observations with a \ac{TTL} of 242.
This example illustrates the inherent uncertainty in mapping \ac{TTL} observations to spoofing sources in absence of ground truth.
In~\S\ref{sec:simulating_attacks}, we evaluate the accuracy of both estimators under realistic routing conditions.
This allows us to verify that the estimators only rarely overestimate the number of spoofing sources.

\section{Results}%
\label{sec:results}

We turn to the results of our measurement and analysis.
We begin by describing the data collected from honeypots~(\S\ref{sec:dataset}).
We then analyze how \ac{TTL} randomization might influence our estimations~(\S\ref{sec:sensitivity_to_ttl_manipulation}).
Next, we apply our estimators to attack data and evaluate their performance in determining the number of spoofing sources~(\S\ref{sec:estimating_spoofing_source_count}).
We further examine the relationship between the scale of an attack and the number of spoofing sources~(\S\ref{sec:attack_scale_and_spoofing_sources}).
Finally, we compare the estimators using simulated attack events~(\S\ref{sec:simulating_attacks}).

\subsection{Dataset}%
\label{sec:dataset}

Our system collects requests to 32~honeypots and groups attacks according to the AmpPot attack definition~(\S\ref{sec:monitoring_attacks}).
The dataset spans two observation periods.
The first period covers October 24, 2024 through February 7, 2025, during which we captured \num{5.49}\,B requests, grouped into \num{843}\,K attacks.
The second period covers August 15, 2025 through February 10, 2026, during which we captured \num{19.3}\,B requests, grouped into \num{2.93}\,M attacks.
The honeypots were continuously active from approximately one month before the start of the first period to the end of the second period.%
	\footnote{Between the observation periods, no data was collected due to an interruption in the measurement pipeline.}
As a result, the second period may capture more activity from attackers reusing previously discovered amplifiers, since the honeypots had been exposed for a longer period~\cite{GriffioenEtAl_ScanTestExecute_2021}.

Both the \ac{TTL} stability calibration~(\S\ref{sec:measuring_ttl_stability}) and attack simulation (\S\ref{sec:simulating_attacks}) were conducted on February 9, 2026.
The stability measurement coincides with the second observation period, but not the first.
Any changes in upstream connectivity in the intervening period may therefore alter the \ac{TTL} variability profile for a honeypot.
Due to this effect, estimates for the second period may be slightly more accurate than those for the first period.
However, our repeated calibration experiment showed that the derived thresholds remained relatively stable over time~(\S\ref{sec:measuring_ttl_stability}).
The calibration and simulation datasets will be made publicly available upon publication of this work.

\subsection{Sensitivity to TTL manipulation}%
\label{sec:sensitivity_to_ttl_manipulation}

To use \ac{TTL} values as an estimator for the number of spoofing sources, we first examined their distributions.
To this end, we analyzed observed \ac{TTL} values in requests that are part of attacks to detect evidence of attackers varying the \ac{TTL} value, as such behavior would skew the outcomes of our estimation algorithms~(\S\ref{sec:computing_an_interval_cover}).

\begin{figure}[tpb]
	\centering
	\includegraphics[width=\linewidth]{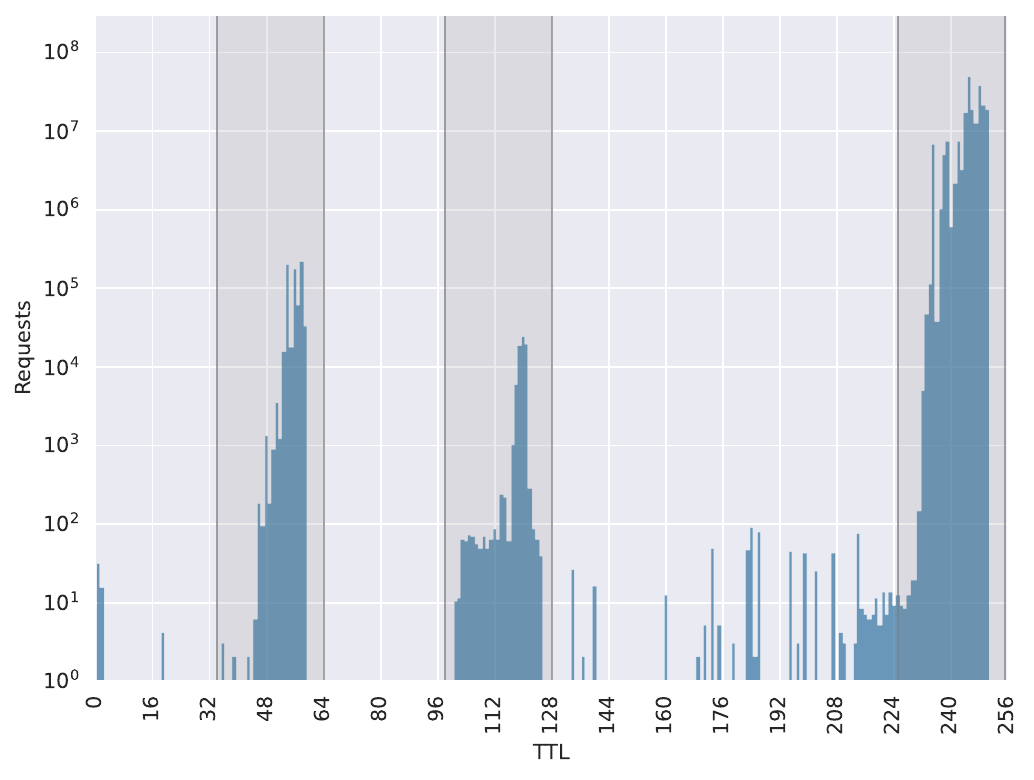}
	\Description{}
	\caption{%
		Histogram of \acs{TTL} values in attacks observed at honeypots in the first observation period.
		Shaded regions indicate the allowed \acs{TTL} ranges, corresponding to at most 30~hops from default initial \ac{TTL} values.
		For visual clarity, we only show the aggregated counts.
	}%
	\label{fig:ttl_counts}
\end{figure}

\begin{figure}[tpb]
	\centering
	\includegraphics[width=\linewidth]{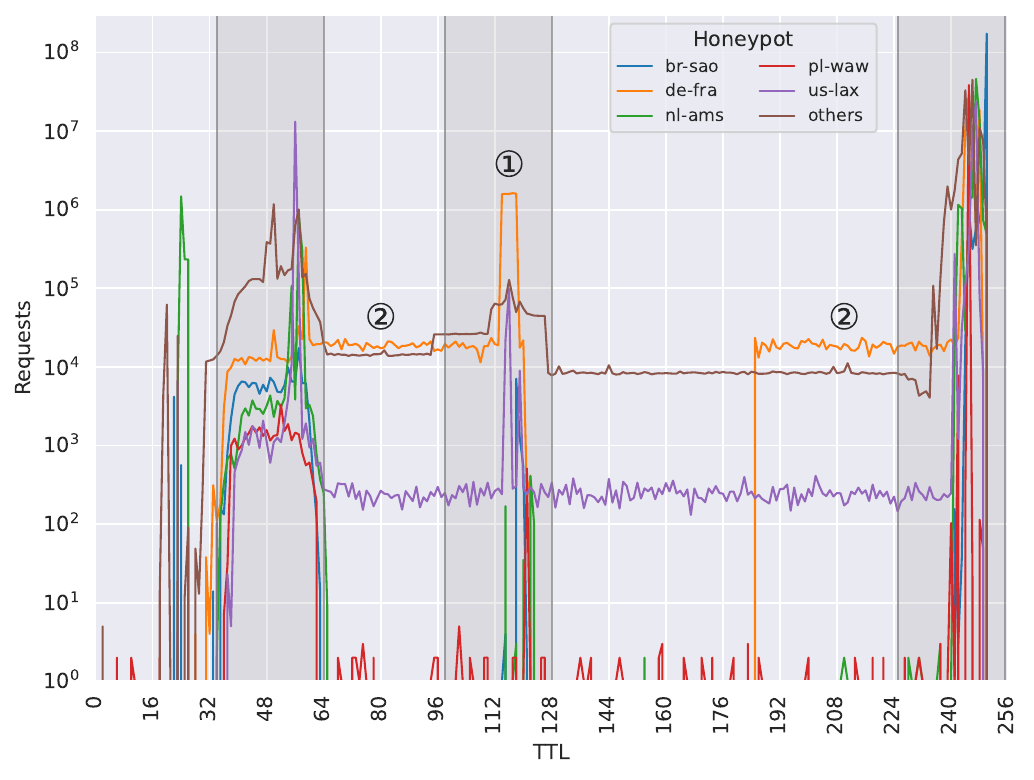}
	\Description{}
	\caption{%
		Histogram of \acs{TTL} values in attacks observed at the five honeypots receiving most traffic in the second observation period.
		Compared to the first observation period, a non-negligible share of traffic is observed with initial values of 64 and 128.
		\circled{1} and \circled{2} indicate attackers varying \acs{TTL} values.
		Shaded regions indicate the allowed \acs{TTL} ranges as in Figure~\ref{fig:ttl_counts}.
	}%
	\label{fig:ttl_top5}
\end{figure}

We first mapped observed \ac{TTL} values to their inferred initial \ac{TTL}.
In the first observation period~(Figure~\ref{fig:ttl_counts}), 99.6\% of requests had an initial \ac{TTL} value of 255.
The behavior changed in the second observation period~(Figure~\ref{fig:ttl_top5}).
Although a majority (93.5\%) of the \ac{TTL} values still started at 255, we saw an increase in the use of lower \ac{TTL} values: 4.4\% of requests had an initial \ac{TTL} of 128, and 2.0\% had an initial \ac{TTL} of 64.
Additionally, we observed greater variation in the \ac{TTL} distributions across our honeypots in the second observation period, possibly due to shifts in attack patterns or increased awareness of our honeypots to attackers.
Figure~\ref{fig:ttl_top5} shows this for the five honeypots receiving the most traffic.

Our estimators are robust to attackers using a random yet single \ac{TTL} value, as this behavior only offsets the observed \ac{TTL} values.
However, attackers that spoof \ac{TTL} values for each request would cause an overestimation of the number of spoofing sources.
The horizontal lines in Figure~\ref{fig:ttl_top5} (\circled{1} and \circled{2}) indicate non-constant \ac{TTL} usage, both observed at \texttt{de-fra}. We further examined these cases.
Case \circled{1} shows a cluster of \ac{NTP} attacks with \ac{TTL} values ranging from 114 to 118, occurring in approximately equal proportions and targeting \acp{AS} in Hong Kong.
Case \circled{2} shows a cluster of attacks that combine \ac{NTP} and the \acf{LDAP}, with \ac{TTL} values uniformly distributed across the ranges 57--120 and 185--245, predominantly targeting European \acp{AS}.
While these cases are readily identifiable due to their uniform \ac{TTL} distributions, other \ac{TTL} manipulation strategies may be more subtle and harder to detect.

To mitigate the impact of \ac{TTL} manipulation and prevent the resulting overestimation, we discarded attacks with requests exceeding 30~hops from the default initial \ac{TTL} values.
In other words, we only retain attacks in which \ac{TTL} values fall within 34--64, 98--128, or 225--255; these are indicated by the shaded regions in Figures~\ref{fig:ttl_counts}~and~\ref{fig:ttl_top5}.
A hop count of 30 corresponds to the default maximum hop count used in the original Unix traceroute implementation~\cite{Jacobson_Traceroute8ManualPage_1989}.
In practice, path lengths are shorter due to increased peering~\cite{DhamdhereDovrolis_InternetFlatModeling_2010} and cloud provider connectivity~\cite{ArnoldEtAl_CloudProviderConnectivity_2020}, as well as our use of 32~globally distributed anycast nodes, which places us closer to the sources of spoofed traffic.
Consistently, no packets had a hop count over~30 in our calibration measurement.

While we received substantial traffic outside these ranges, the impact on the dataset was small: we discarded \num{29}~($<0.01\%$) attacks comprising \num{7.93}\,M requests in the first observation period and \num{42.4}\,K~(1.45\%) attacks comprising \num{407}\,M requests in the second.
We note that this imposes a limitation: our filtering procedure cannot detect attacks that manipulate \ac{TTL} values exclusively within the 30-hop ranges.

\subsection{Estimating spoofing source count}%
\label{sec:estimating_spoofing_source_count}

\begin{figure}[tpb]
	\centering
	\begin{subfigure}[t]{\linewidth}
		\centering
		\includegraphics[width=\linewidth]{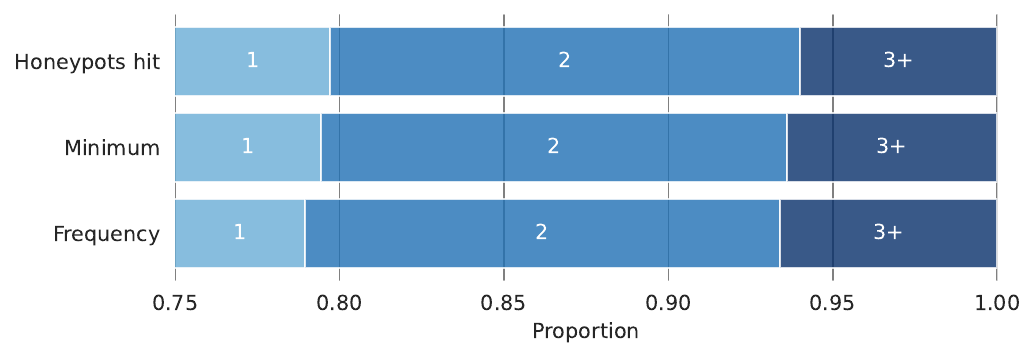}
		\Description{}
		\caption{First observation period}%
		\label{fig:origin_count_obsper1}
	\end{subfigure}%
	\par\medskip
	\begin{subfigure}[t]{\linewidth}
		\centering
		\includegraphics[width=\linewidth]{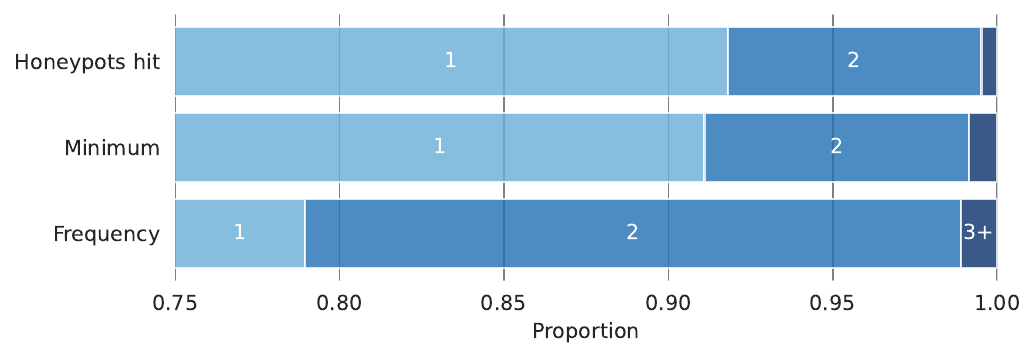}
		\Description{}
		\caption{Second observation period}%
		\label{fig:origin_count_obsper2}
	\end{subfigure}%
	\caption{%
		Number of spoofing sources by estimator: distinct \emph{honeypots hit}, \emph{minimum} cover, and \emph{frequency}-based.
		The majority of attacks are estimated to stem from a single spoofing source.
	}
	\label{fig:origin_count}
\end{figure}

We applied our estimators to all attacks reaching the anycast address of our honeypots recorded in both observation periods, after the filtering procedure outlined in the previous section.
As explained in~\S\ref{sec:monitoring_attacks}, we restrict our analysis to anycast because it distributes spoofed attack traffic over multiple honeypots, which helps distinguish traffic from different spoofing sources.

Although the minimum cover estimator is already conservative, it estimated slightly more sources than the naive approach of only counting the honeypots hit.
The frequency-based estimator produced the highest estimates overall and identified that 21.0\% of attacks originated from more than one spoofing source.
Figure~\ref{fig:origin_count} shows the number of estimated spoofing sources across estimators and observation periods.

In the first observation period, the estimated number of spoofing sources did not vary substantially across estimators.
However, in the second period, the frequency-based estimator identified 2.47$\times$ as many attacks from two sources compared to the minimum cover estimator.
To explain this difference, we examined all cases where the minimum cover estimator identified only a single spoofing source but the frequency-based estimator identified two.
A large majority~(97.7\%) of these attacks contacted only the \texttt{de-fra} honeypot and targeted \acp{AS} in Brazil using \ac{LDAP} between October~28 and December~8, 2025, spanning 291~prefixes across 48~\acp{AS}.
These attacks often exhibited exactly three adjacent \ac{TTL} values, which makes it ambiguous whether the observations originate from a single source with path variability or from multiple sources with similar path lengths to the honeypot.
Because \texttt{de-fra} uses a \ac{TTL} stability threshold $j_h = 1$, both interpretations can yield different estimations: the minimum cover estimator groups all values into a single interval centered at the midpoint, while the frequency-based estimator centers intervals on the most frequent \ac{TTL} value, which may result in two intervals.
This cluster alone accounts for 91.6\% of the cases where the minimum cover estimator identified one source, whereas the frequency-based estimator identified two.

Focusing on the long tail of attacks, we observed two attacks on the same day in the first observation period that shared the highest number of estimated sources for both the minimum cover~(24) and the frequency-based~(31) estimator.
Both used the same three honeypots in Brazil, France, and the Netherlands.
We also saw attacks originating from a wide geographical distribution.
In particular, two attacks in the first period, 17~days apart but targeting the same host, reached 11~honeypots across Africa, the Americas, Asia, and Europe.
The \ac{TTL} values in these attacks were more concentrated, with both estimators inferring 13~sources in each case.
Finally, in the second period, one attack ranked highest across both estimators, hitting 29~honeypots and totaling 215 or 272 sources, as estimated by the minimum cover and frequency-based estimators, respectively.
This attack, lasting 21~hours, targeted a hosting provider in the United States and originated from almost all regions worldwide.

For comparison, Krämer et al.~\cite{KramerEtAl_AmpPotMonitoringDefending_2015} used 21~unicast honeypots and applied a static limit on the number of distinct \ac{TTL} values per attack to estimate that only 3.7\% of attacks stem from multiple sources.
The higher share of multi-source attacks in our results may suggest that spoofing infrastructure has become more widely distributed over the past decade.
More generally, such estimates depend strongly on the underlying measurement methodology and the assumptions used to model \ac{TTL} variability.
Our approach therefore focuses on deriving conservative lower-bound estimates using per-honeypot calibration, rather than attempting to infer the exact number of spoofing sources.

\subsection{Attack scale and spoofing sources}%
\label{sec:attack_scale_and_spoofing_sources}

\begin{figure}[tpb]
	\centering
	\includegraphics[width=\linewidth]{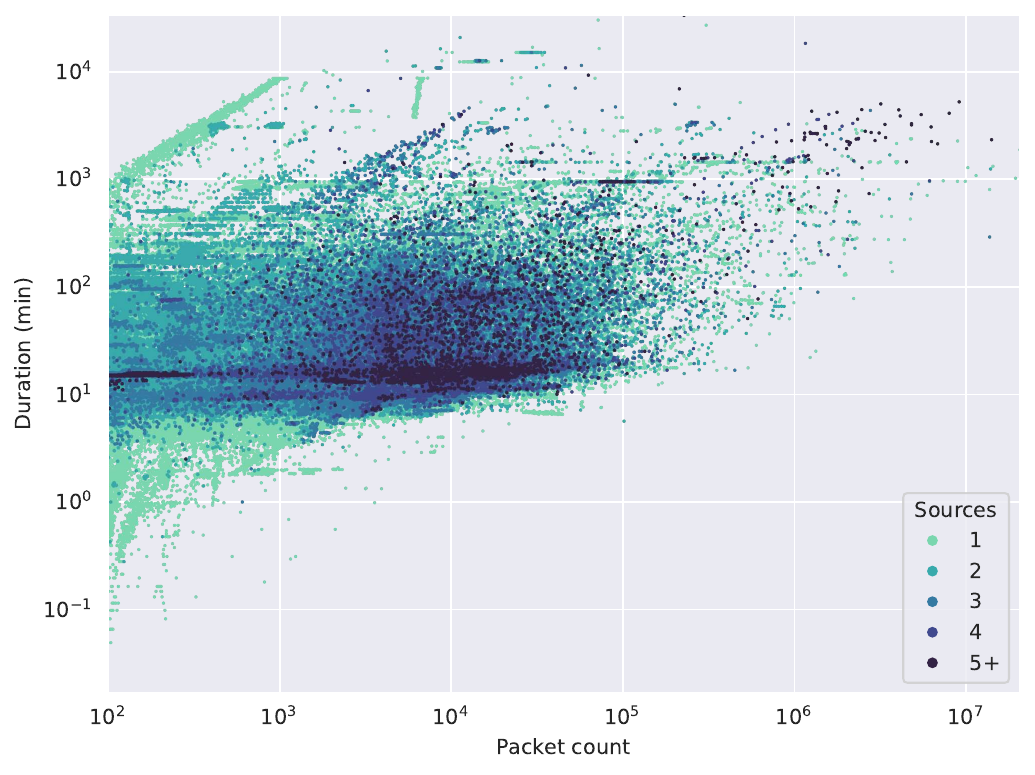}
	\Description{}
	\caption{%
		Attack scale, measured by duration and packet count, and the number of spoofing sources estimated using the frequency-based estimator in the first observation period.
		Attack duration was weakly positively correlated with the number of spoofing sources, while packet count showed no meaningful relationship.
	}%
	\label{fig:attack_size_origins}
\end{figure}

We investigated whether large-scale attacks involve more spoofing sources than smaller ones.
To this end, we examined the relationship between the number of spoofing sources estimated by the frequency-based estimator and two indicators of attack scale:
	packet count\footnote{After sampling described in~\S\ref{sec:rate_limiting_and_sampling}.}
	and duration.

Figure~\ref{fig:attack_size_origins} presents attack scale, measured by duration and packet count, versus the number of estimated spoofing sources.
Due to the skew toward single-source attacks, we plotted attacks in ascending order of source count, ensuring that attacks with more sources are drawn on top.
The figure shows a wide range of packet counts and durations, where longer attacks tended to have higher packet counts.
However, it also shows multiple-source attacks occurred at both small and large scales.

To quantify these relationships, we computed Spearman correlations between spoofing source count and both packet count and attack duration.
In the first observation period, the correlation between packet count and spoofing source count was negligible~($\rho=0.020$), while the correlation with attack duration was weak~($\rho=0.221$).
Due to the large number of attacks included in the analysis, all $p$-values were below~0.001.
Results for the second observation period were similar.

Overall, these results suggest that the estimated number of spoofing sources is largely independent of attack scale.
This weak association may reflect that prolonged attacks maintain or rotate multiple spoofing sources over time, whereas high packet rates can also be generated by just a single spoofing source.

\subsection{Simulating attacks}%
\label{sec:simulating_attacks}

\begin{table}[tpb]
	\centering
	\Description{}
	\caption{%
		Performance of the minimum cover and frequency-based estimators, showing absolute and relative counts at 1, 2, and 3 simulated sources.
		Entries below the diagonal are underestimations, and above the diagonal are overestimations.
	}%
	\label{tab:simulation_results}
	\small
	\setlength{\tabcolsep}{3pt}
	\begin{tabular*}{\linewidth}{
		@{\extracolsep{\fill}}
		l
		c
		S[table-format=5]@{\enspace}r
		S[table-format=5]@{\enspace}r
		S[table-format=4]@{\enspace}r
	}
		\toprule
			\multirow{2}{*}[-2pt]{Estimator} &
			\multirow[b]{2}{*}[-2pt]{\shortstack{Sources\\[0.3ex] simulated}} &
			\multicolumn{6}{c}{Sources estimated} \\
			\cmidrule(l{2pt}r{2pt}){3-8}
			 & & \multicolumn{2}{c}{1} & \multicolumn{2}{c}{2} & \multicolumn{2}{c}{3} \\
		\midrule
			          & 1 &   441 & (100.0\%) &     0 &  (0.0\%) &    0 & (0.0\%) \\
			  Minimum & 2 &  9194 &  (75.0\%) &  3058 & (25.0\%) &    0 & (0.0\%) \\
			          & 3 & 42618 &  (61.2\%) & 26235 & (37.7\%) &  766 & (1.1\%) \\
		\midrule
			          & 1 &   438 &  (99.3\%) &     3 &  (0.7\%) &    0 & (0.0\%) \\
			Frequency & 2 &  6637 &  (54.2\%) &  5613 & (45.8\%) &    2 & (0.0\%) \\
			          & 3 & 28726 &  (41.3\%) & 36512 & (52.4\%) & 4380 & (6.3\%) \\
		\bottomrule

	\end{tabular*}
\end{table}

We conducted an additional experiment to assess the risk of source overestimation under realistic routing conditions, which would inflate estimates of attackers' resources and capabilities.
This experiment is intended to test whether the assumptions underlying the algorithms hold and deliberately does not replicate all characteristics of a \ac{DDoS} attack, as doing so would raise ethical concerns.
As a result of not sending packets at rates reflecting actual \ac{DDoS} attacks, we do not trigger congestion-driven load balancing, which may slightly increase the observed thresholds.
This, in turn, may lead to a lower estimated number of spoofing sources in an actual attack scenario.

We used a subset of RIPE Atlas probes located in networks without ingress filtering, as defined in \acs{BCP}~38~\cite{rfc2827}, identified using recent CAIDA Spoofer measurements~\cite{CenterforAppliedInternetDataAnalysisCAIDA_Spoofer_2026}.
These networks are arguably more representative of permissive \acp{AS}, from which spoofed attack traffic could originate.
From this set, we restrict our selection to networks tested within the preceding 30~days, yielding 469~probes.
From each probe, we sent 55~packets per probe over 10~minutes, the maximum reliable rate.
We observed responses from 429~probes and excluded 5 with fewer than 50~packets.

We then evaluated how well the two \acs{TTL}-based estimators separate traffic from simulated sources.
For each honeypot, we combined traffic from one, two, or three probes, sampling up to \num{10000} combinations per setting.
Table~\ref{tab:simulation_results} summarizes the results.
For instance, when simulating two sources, the frequency-based estimator inferred three sources in only two simulations.
Across all simulations, only one simulation yielded four estimated sources (not shown).

Both estimators rarely overestimated, indicating that they provide reliable lower bounds on the number of sources.
The minimum cover estimator never overestimated, while the frequency-based estimator did so in only six out of \num{82312} simulations~(0.007\%).
Nevertheless, the frequency-based estimator yields tighter bounds, as reflected in more exact matches and more frequent estimates of two sources when simulating three.

\section{Discussion}%
\label{sec:discussion}

We have demonstrated how the topological diversity inherent in anycast deployments can be harnessed to detect \ac{DRDoS} attacks globally and estimate a lower bound on the number of spoofing sources involved.
To this end, we introduced two estimation algorithms based on observed \ac{TTL} values.
The frequency-based estimator may more realistically distinguish between spoofing sources, at a small risk of overestimation.
This trade-off may be appropriate when a tighter bound is preferred over strictly avoiding overestimation, such as for planning defensive strategies and allocating resources by \acp{CDN}, \acp{DPS}, and \acp{SOC}.

Beyond estimation, the topological distribution of spoofers provides additional insight into attacker behavior.
Although the number of \ac{TTL}/honeypot combinations is likely not large enough to uniquely identify the routed prefix or \ac{AS}, this information can still contribute to building a more complete profile of threat actors when combined with other fingerprinting mechanisms, such as scanner infrastructure~\cite{KruppEtAl_IdentifyingScanAttack_2016}, domain names~\cite{NawrockiEtAl_FarSideDNS_2021}, or other application-layer features.
However, we have seen attackers randomizing initial \acs{TTL} values to thwart analysis.
Developing methods robust to \acs{TTL}-based evasion is therefore an important direction for future work.

Our results indicate that spoofed requests fueling \ac{DRDoS} attacks tend to originate from a small set of networks, rather than tens or hundreds of them.
One possible explanation for this limited spoofing footprint is that attackers minimize their exposed infrastructure by employing only the necessary number of spoofing sources to ensure attack success while avoiding unnecessary detection risks.
It also supports the viability of attack attribution by manipulating traffic paths, which we leave as future work.

\subsection{Threshold parameter selection}%
\label{sec:threshold_parameter_selection}

To derive a threshold for path fluctuations, we calculated the 95\textsuperscript{th} percentile over the ranges of observed \ac{TTL} values.
Selecting a higher percentile would raise the threshold values, thereby reducing the estimated number of spoofing sources.
For example, choosing the 99\textsuperscript{th} percentile increases the threshold for 25 of the 32 honeypots.
In contrast, a lower percentile would lower the thresholds, increasing the risk of overestimating the number of spoofing sources.
Since \ac{TTL} values are integers, dividing by two can yield either an integer or a half-integer.
We round up rather than down to obtain a more liberal (i.e.,~wider) threshold.
A wider threshold reduces the risk of overestimation, which we aim to avoid, at the cost of allowing more underestimations.
In~\S\ref{sec:simulating_attacks}, we showed that the empirically established thresholds exhibit negligible overestimation while making the estimated lower bound more accurate (i.e.,~closer to the true value than the minimum cover estimator).

\section{Conclusion}%
\label{sec:conclusion}

In this work, we presented a methodology to estimate a lower bound on the number of spoofing sources participating in \ac{DRDoS} attacks.
We deployed amplification honeypots in an anycast environment to attract localized attack activity and collect associated \ac{TTL} values.
We calibrated our estimators using empirically derived \ac{TTL} stability thresholds and evaluated their performance under realistic routing conditions, showing that avoiding overestimations can be traded off for tighter lower bounds on the number of spoofing sources.

Our results show that 21.0\% of attacks involve multiple spoofing sources, a much higher share than reported in earlier work.
This shift could reflect changes in the attack landscape, with attackers increasingly employing multiple sources to improve attack resilience and complicate defenses.

Our findings indicate that assuming a single spoofing source is not generally justified, as a non-negligible share of attacks distribute spoofing sources across multiple networks.
This property should be reflected in the design of future attribution techniques.

\balance
\bibliographystyle{ACM-Reference-Format}
\bibliography{references.bib}

\appendix

\section{Ethical considerations}%
\label{sec:ethical_considerations}

\subsection{Amplification honeypots}%
\label{sec:amplification_honeypots}

In operating amplification honeypots, it is important to carefully consider any potential harm they may cause.
Prior to the deployment of the honeypots, the study protocol was submitted for review and received approval from the \ac{IRB} to ensure compliance with ethical standards.
We adhered to best practices and followed the default rate limit for AmpPot: our honeypots responded to the first three probes from a given client \ac{IP} address and its covering /24~prefix.
Beyond that, responses were limited to one per client /24~prefix per minute, for a duration of one hour, to minimize network load and reduce the risk of abuse.
Responding at such a conservative rate makes it highly unlikely that our honeypots could cause any harm or disruption.
Furthermore, to prevent misuse and avoid aiding attackers, we do not publish any honeypot data.

\subsection{RIPE Atlas probing}%
\label{sec:ripe_atlas_probing}

We performed two experiments in which we probed our testbed from RIPE Atlas to calibrate~(\S\ref{sec:measuring_ttl_stability}) and simulate~(\S\ref{sec:simulating_attacks}) the estimators introduced in this paper.
Probing at scale poses two risks: overloading the Atlas platform and triggering \ac{DDoS} alarms on the cloud provider supporting our testbed.
In the calibration experiment, we probed from a maximum of \num{15000} Atlas probes.
We spread the probing over one day, resulting in an aggregate rate of 8.33~\ac{pps} across all Atlas probes and a negligible arrival rate of approximately 0.26~\ac{pps} per testbed node.
In the simulation experiment, we sent traffic from fewer sources~(469) but within a shorter time span of 10~minutes.
Here, we probed at an aggregate rate of 43.0~\ac{pps} with an expected arrival rate of 1.34~\ac{pps} per testbed node.
For both experiments, we scheduled only a single measurement with the platform.
At these modest rates, it is highly unlikely to overload systems or trigger alarms.

\section{Detailed calibration results}%
\label{sec:detailed_calibration_results}

\begin{figure*}[tpb]
	\centering
	\captionsetup{width=.8\linewidth}
	\includegraphics[width=.8\linewidth]{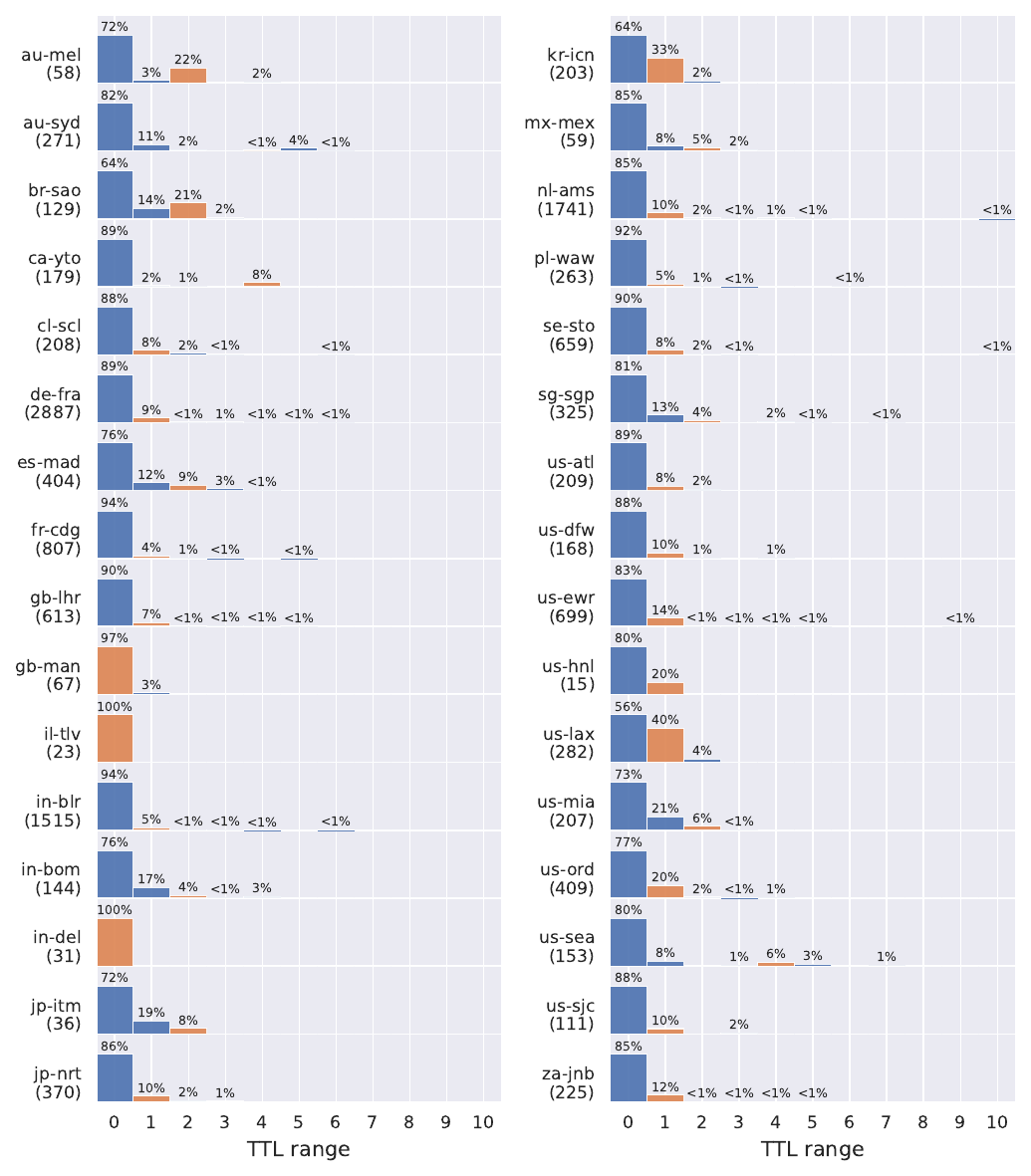}
	\Description{}
	\caption{%
		Distribution of \ac{TTL} ranges~$\delta_{h,a}$ ($x$-axis) over Atlas probes~$a$ for each honeypot~$h$ ($y$-axis).
		Heights are normalized across honeypots.
		Numbers in parentheses indicate the number of unique probes observed by the honeypot.
		The orange bars denote the 95\textsuperscript{th} percentile.
		\texttt{il-tlv} and \texttt{in-del} exhibited no variation in \ac{TTL} ranges.
		}
	\label{fig:ttl_range_distr}
\end{figure*}

\end{document}